\documentclass[letterpaper,conf]{IEEEtran}
\usepackage[T1]{fontenc}
\usepackage[latin9]{inputenc}
\usepackage{color}
\usepackage{units}
\usepackage{amsmath}
\usepackage{amsthm}
\usepackage{amssymb}
\usepackage{graphicx}

\makeatletter

\pdfpageheight\paperheight
\pdfpagewidth\paperwidth

\providecommand{\tabularnewline}{\\}

\theoremstyle{plain}
\newtheorem{thm}{\protect\theoremname}
\theoremstyle{plain}
\newtheorem{lem}[thm]{\protect\lemmaname}

\usepackage{cite}
\usepackage{amsmath,amssymb,amsfonts}
\usepackage{algorithmic}
\usepackage{graphicx}
\usepackage{textcomp}
\usepackage{xcolor}
\usepackage{url}

\usepackage{mathtools}
\allowdisplaybreaks

\usepackage[margin=0.8in]{geometry}
\usepackage{algorithm}
\usepackage{algorithmic}
\usepackage{comment}

\makeatother
\providecommand{\lemmaname}{Lemma}

\providecommand{\theoremname}{Theorem}

\makeatother

\providecommand{\lemmaname}{Lemma}
\providecommand{\theoremname}{Theorem}

\begin{document}

\title{Towards Power-Efficient Aerial Communications via Dynamic Multi-UAV Cooperation}

\author{\IEEEauthorblockN{Lin Xiang\IEEEauthorrefmark{1}, Lei Lei\IEEEauthorrefmark{1}, Symeon
Chatzinotas\IEEEauthorrefmark{1}, Bj\"{o}rn Ottersten\IEEEauthorrefmark{1}, and Robert Schober\IEEEauthorrefmark{2} \\} \IEEEauthorblockA{\IEEEauthorrefmark{1}Interdisciplinary Center for Security, Reliability and Trust (SnT), University of Luxembourg \\}\IEEEauthorblockA{\IEEEauthorrefmark{2}Institute for Digital Communications, Friedrich-Alexander University of Erlangen-Nuremberg \vspace{-1cm}}}
\maketitle
\begin{abstract}
Aerial base stations (BSs) attached to unmanned aerial vehicles (UAVs) constitute a new paradigm for next-generation cellular communications. However, the flight range and communication capacity of aerial BSs
are usually limited due to the UAVs' size, weight, and power (SWAP) constraints. To address this challenge, in this paper, we consider \emph{dynamic} cooperative transmission among multiple aerial BSs for power-efficient aerial communications. Thereby, a central controller intelligently selects the aerial BSs navigating in the air for cooperation. Consequently, the large virtual array of moving antennas formed by the cooperating aerial BSs can be exploited for low-power information transmission and navigation, taking into account the channel conditions, energy availability, and user demands. Considering both the fronthauling and the data transmission links, we jointly optimize the trajectories, cooperation decisions, and transmit beamformers of the aerial BSs
for minimization of the weighted sum of the power consumptions required by all BSs. Since obtaining the global optimal solution of the formulated problem is difficult, we propose a low-complexity iterative algorithm
that can efficiently find a Karush-Kuhn-Tucker (KKT) solution to the problem. Simulation results show that, compared with several baseline schemes, dynamic multi-UAV cooperation can significantly reduce the communication and navigation powers of the UAVs to overcome the SWAP limitations, while requiring only a small increase of the transmit power over the fronthauling links. 
\end{abstract}

\vspace{-.4cm}

\section{Introduction}

Exploiting unmanned aerial vehicles (UAVs) or drones as aerial base
stations (BSs) for enhanced cellular communication has recently attracted
significant interest \cite{Zeng16UAV,Halim19}. Unlike terrestrial
BSs whose communication with ground users is usually subject to non-line-of-sight
(NLoS) channels, aerial BSs can proactively seek line-of-sight (LoS)
connections with ground users to facilitate favorable signal propagation.
Moreover, the deployment of UAVs can be adapted on demand to the spatial
and temporal distributions of the cellular users under both normal
and contingency conditions. Yet, aerial BSs are usually constrained
in size, weight, and power supply (SWAP) and have only limited flight
range and communication capabilities \cite{Tran19}. Therefore, improving
the navigation and communication performance of aerial BSs within
the SWAP limits is a crucial research challenge. 

A promising approach is to employ an array of networked UAVs, whereby
the existing aerial communication schemes for networked UAVs can be
classified into\emph{ non-cooperative }\cite{QQWu18MultiUAV,Zhang19Access}
and \emph{cooperative} \cite{Liu18UAVCoMP,Liu18multibeam} schemes.
For the non-cooperative schemes, the navigation/communication tasks
are divided among the UAVs across time and space, which leads to low
payload and low communication overheads per UAV \cite{QQWu18MultiUAV}.
However, these schemes require orthogonal spectrum allocation for
the UAVs and coexisting terrestrial BSs/users, leading to a low spectrum
utilization. Otherwise, LoS co-channel interferers may severely degrade
the reliability of aerial communications \cite{UAV-LTE,Yajna18IM}. 

To boost network capacity, multi-UAV cooperation with full frequency
reuse, akin to the multi-cell cooperation paradigm in cellular communications
\cite{Fettweis11CoMP}, has been proposed. In \cite{Liu18UAVCoMP},
multiple UAVs hovering in the air are utilized as aerial remote radio
heads (RRHs) to communicate with ground users in the uplink, and the
signals received at each UAV are forwarded to a central processor
for joint decoding. The authors investigate the optimal placement
and movement of the UAVs for maximization of the minimal achievable
rate of the users \cite{Liu18UAVCoMP}. A BS cooperation scheme for
canceling the interference caused by multi-antenna UAVs is proposed
in \cite{Liu18multibeam}. In particular, each BS forwards its decoded
message(s) to the other BSs via backhaul links, which are then exploited
for interference cancellation. The authors in \cite{Liu18multibeam}
investigate optimal beamforming for maximization of the sum rate for
UAVs hovering at fixed positions.

The aforementioned works \cite{QQWu18MultiUAV,Zhang19Access,Liu18UAVCoMP,Liu18multibeam}
assume the non-cooperation and cooperation among UAVs to be \emph{fixed}
over time and space. However, due to the UAVs' mobility and the heterogeneity
of the terrain features, the signal and interference powers in aerial
networks may vary significantly along the UAVs' flying trajectories,
which cannot be properly exploited with the existing static schemes.
To further unlock the potential of networked UAVs, in this paper,
we introduce the new concept of \emph{dynamic} multi-UAV cooperation.
Thereby, the UAVs are intelligently selected for cooperation with
other UAVs, taking into account their positions/trajectories, channel
and energy conditions, and the users' demands. By exploiting the resulting
large virtual array of moving antennas for cooperative data transmission,
the UAVs' mechanical navigation\footnote{For example, a navigating UAV can seek LoS/NLoS signal propagation
paths and/or move close to the desired users and/or away from the
interferers. } and cooperative beamforming provide additional spatial degrees of
freedom for facilitating power-efficient aerial communications.

To maximize the benefits of dynamic multi-UAV cooperation within the
SWAP constraints, we jointly optimize the UAVs' trajectories, cooperation
decisions, and cooperative beamformers for minimization of the weighted
sum of the BSs' power consumptions while guaranteeing the quality
of service (QoS) of the users and safe navigation of the UAVs. The
formulated problem is a mixed-integer non-convex program and finding
the global optimal solution is generally NP-hard. To tackle this issue,
we exploit the underlying difference of convex (DC) program structure
and propose a low-complexity suboptimal scheme based on binary approximation
and the convex-concave procedure (CCP) \cite{Lipp16CCP}. Under mild
conditions, the solution to the joint optimization problem found by
the proposed algorithm fulfills the Karush\textendash Kuhn\textendash Tucker
(KKT) optimality conditions of the original non-convex problem. The
contributions of this paper are summarized as follows:
\begin{itemize}
\item We propose dynamic multi-UAV cooperation, where each UAV can intelligently
cooperate with other UAVs during navigation, to enable power-efficient
aerial communication. 
\item We develop a low-complexity algorithm for joint optimization of the
UAVs' cooperation, the fronthauling and data transmission, and the
UAVs' trajectories to minimize the weighted sum of the BSs' power
consumptions required for communication and navigation.
\item Our simulation results show that dynamic multi-UAV cooperation can
significantly improve the power efficiency of UAV communication and
navigation despite the UAVs' SWAP constraints. 
\end{itemize}
\emph{Notations:} Throughout this paper, $\mathbb{R}$, $\mathbb{R}_{+}$,
and $\mathbb{C}$ denote the sets of real, non-negative real, and
complex numbers, respectively. $\mathbb{C}^{N}$ and $\mathbb{C}^{N\times M}$
are the sets of complex $N\times1$ vectors and $N\times M$ matrices,
respectively. $\mathbf{I}_{N}$ is the $N\times N$ identity matrix.
$\Re\left\{ \mathbf{z}\right\} $ and $\Im\left\{ \mathbf{z}\right\} $
denote the real and imaginary parts of complex-valued vector $\mathbf{z}\in\mathbb{C}^{N}$,
respectively. $(\cdot)^{\mathrm{T}}$, $(\cdot)^{\mathrm{H}}$, $\mathrm{tr}(\cdot)$,
and $\mathrm{rank}(\cdot)$ are the transpose, complex conjugate transpose,
trace, and rank operators, respectively. $\left|\cdot\right|$, $\left\Vert \cdot\right\Vert $,
and $\left\Vert \cdot\right\Vert _{F}$ denote the absolute value
of a scalar, the $\ell_{2}$-norm of a vector, and the Frobenius-norm
of a matrix, respectively. $\mathbf{x}\preceq\mathbf{y}$ ($\mathbf{x}\succeq\mathbf{y}$)
means that vector $\mathbf{x}$ is element-wise smaller (greater)
than or equal to vector $\mathbf{y}$. $\mathcal{CN}\left(\mu,\sigma^{2}\right)$
represents the complex Gaussian distribution with mean $\mu$ and
variance $\sigma^{2}$. Finally, $\nabla f(\mathbf{x})$ is the gradient
of function $f(\mathbf{x})$ with respect to $\mathbf{x}$.

\vspace{-.4cm}

\section{System Model}

\begin{figure}[t]
\centering\includegraphics[scale=0.52]{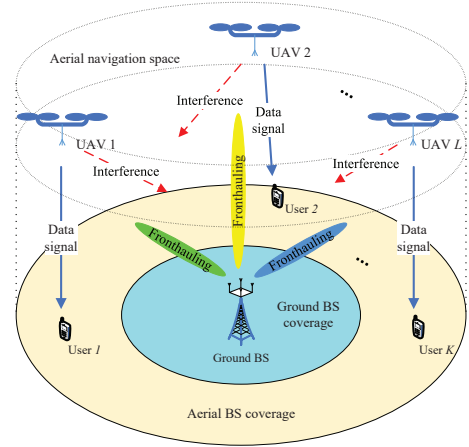}

\caption{\label{fig1}Illustration of multi-UAV assisted downlink communication
for users whose connections to the ground BS are blocked. }

\vspace{-0.5cm}
\end{figure}

We assume that $L$ UAVs, each mounted with a cellular transceiver
(aerial BS), are deployed for providing downlink communications to
$K$ ground users, see Figure \ref{fig1}. Let $\mathcal{L}\triangleq\left\{ 1,\ldots,L\right\} $
and $\mathcal{K}\triangleq\left\{ 1,\ldots,K\right\} $ denote the
index sets of the UAVs and the users, respectively. The UAVs employ
wireless fronthauling by connecting to a remote ground BS. The ground
BS and each UAV are equipped with $N\ge1$ and $M\ge1$ antennas,
respectively, whereas the users are single-antenna devices. We assume
$N\ge L$ and $LM\ge K$ to ensure a feasible problem formulation.
Moreover, due to large propagation distances and potential blockages,
the users of interest cannot establish a direct connection to the
ground BS. Therefore, information transmission to these users comprises
(i) fronthauling from the ground BS to the UAVs and (ii) data transmission
from the UAVs to the users. We consider a time-slotted system. Each
time slot is divided into two intervals of equal duration, where fronthauling
and data transmission are performed the first and the second interval
of each time slot, respectively\footnote{The system model is also applicable for fronthauling and data transmission
over orthogonal frequency bands, e.g., over mmWave and sub-6 GHz bands,
respectively.}. 

The UAVs may navigate within a given aerial space to facilitate communication
with the ground BS and the users. However, depending on the UAVs'
positions, the channel conditions for aerial communications, including
the LoS/NLoS propagation paths and the interference caused by multi-UAV
fronthauling and data transmission, may vary significantly. Hence,
dynamic cooperation among the UAVs is desirable to coordinate information
transmission, interference mitigation, and navigation in real-time.
In the following, we first investigate the underlying aerial-to-ground
channels and then present the aerial communication design tailored
to the channel characteristics. 

\vspace{-.4cm}

\subsection{Channel Modeling}

Let $\mathbf{d}_{k}^{\mathrm{UE}}\triangleq\left[d_{k,x}^{\mathrm{UE}},d_{k,y}^{\mathrm{UE}},d_{k,z}^{\mathrm{UE}}\right]^{\mathrm{T}}\in\mathbb{R}^{3}$
and $\mathbf{d}^{\mathrm{BS}}\triangleq\left[d_{x}^{\mathrm{BS}},d_{y}^{\mathrm{BS}},d_{z}^{\mathrm{BS}}\right]\in\mathbb{R}^{3}$
denote the fixed positions of user $k\in\mathcal{K}$ and the ground
BS, respectively. Furthermore, $\mathbf{d}_{l,t}\triangleq\left[d_{l,x,t},d_{l,y,t},d_{l,z,t}\right]^{\mathrm{T}}\in\mathbb{R}^{3}$
denotes the position of UAV $l\in\mathcal{L}$ at time $t$. The distance
between UAV $l$ and user $k$ at time $t$ is thus given by 
\[
\begin{alignedat}{1} & d_{l,k,t}\triangleq\left\Vert \mathbf{d}_{l,t}-\mathbf{d}_{k}^{\mathrm{UE}}\right\Vert \\
 & =\sqrt{(d_{l,x,t}-d_{k,x}^{\mathrm{UE}})^{2}\!+\!(d_{l,y,t}-d_{k,y}^{\mathrm{UE}})^{2}\!+\!(d_{l,z,t}-d_{k,z}^{\mathrm{UE}})^{2}}.
\end{alignedat}
\]
Likewise, the distance between UAV $l$ and the ground BS at time
$t$ is $d_{\mathrm{F},l,t}\triangleq\left\Vert \mathbf{d}_{l,t}-\mathbf{d}^{\mathrm{BS}}\right\Vert $.
Due to ground reflections and scattering, data signals transmitted
over the UAV-to-user channel may undergo both LoS and NLoS propagation.
Let $\mathbf{h}_{l,k,t}\in\mathbb{C}^{M\times1}$ be the channel gain
vector between UAV $l$ and user $k$ at time $t$. We assume $\mathbf{h}_{l,k,t}=\sqrt{A_{l,k}d_{l,k,t}^{-\alpha_{l,k}}}\mathbf{g}_{l,k,t}$,
where $A_{l,k}d_{l,k,t}^{-\alpha_{l,k}}$ and $\mathbf{g}_{l,k,t}\in\mathbb{C}^{M}$
capture the propagation path loss and the channel gains due to multipath
fading, respectively. $\alpha_{l,k}$ is the path loss exponent of
the channel between UAV $l$ and user $k$, and $A_{l,k}$ is a constant
accounting for the antenna gains. 

On the other hand, as both antenna arrays are elevated, the ground
BS-to-UAV fronthauling channels are usually dominated by LoS propagation.
Let $\mathbf{H}_{\mathrm{F},l,t}\in\mathbb{C}^{N\times M}$ be the
channel matrix between the ground BS and UAV $l$ at time $t$. Without
loss of generality, we assume $\mathbf{H}_{\mathrm{F},l,t}=\sqrt{A_{\mathrm{F},l}d_{\mathrm{F},l,t}^{-\alpha_{\mathrm{F},l}}}\mathbf{G}_{\mathrm{F},l,t}$,
where $\alpha_{\mathrm{F},l}$ and $A_{\mathrm{F},l}$ denote the
path loss exponent and the antenna gains, respectively. The channel
gain matrix at time $t$, $\mathbf{G}_{\mathrm{F},l,t}\in\mathbb{C}^{N\times M}$,
typically has a low rank due to LoS propagation \cite[Ch. 7.2.3]{Tse05Fundamentals}.
Hence, we approximate $\mathbf{G}_{\mathrm{F},l,t}$ as $\mathbf{G}_{\mathrm{F},l,t}\approx\mathbf{g}_{\mathrm{F},l,t}^{\mathrm{tx}}(\mathbf{g}_{\mathrm{F},l,t}^{\mathrm{rx}})^{\mathrm{H}}$
with $\mathbf{g}_{\mathrm{F},l,t}^{\mathrm{tx}}\in\mathbb{C}^{N}$
and $\mathbf{g}_{\mathrm{F},l,t}^{\mathrm{rx}}\in\mathbb{C}^{M}$. 

Throughout this paper, we consider block fading channels, where $\mathbf{g}_{l,k,t}$
and $\mathbf{G}_{\mathrm{F},l,t}$ remain constant over a block of
$T$ time slots but vary independently from one block to the next.
This is because the flight speed of UAVs is usually low and hence,
the duration of a time slot is much smaller than the coherence time
of the channel. In the following, we present the system model and
problem formulation for one block with the time slots indexed by set
$\mathcal{T}\triangleq\left\{ 1,\ldots,T\right\} $. For convenience,
we rewrite $\mathbf{g}_{l,k,t}$ and $\mathbf{G}_{\mathrm{F},l,t}$
as $\mathbf{g}_{l,k}$ and $\mathbf{G}_{\mathrm{F},l}$, respectively.

\vspace{-.4cm}

\subsection{Dynamic UAV Cooperation for Data Transmission}

We assume that a central controller (e.g. located at the ground BS,
see also Section \ref{sec3}) intelligently selects the UAVs for cooperation
according to the UAVs' positions and battery status, the channel state,
and the users' QoS requirement for serving the users. Let $q_{l,k}=1$
if UAV $l\in\mathcal{L}$ serves user $k\in\mathcal{K}$, and $q_{l,k}=0$
otherwise. The UAVs indexed by set $\mathcal{L}_{k}\triangleq\left\{ l\mid q_{l,k}=1\right\} $
employ cooperative beamforming for data transmission to user $k$.
For a low-complexity implementation, the cooperation decisions $\left\{ q_{l,k}\right\} $
are fixed within $\mathcal{T}$. We assume that all UAVs are synchronized\footnote{The UAVs are usually equipped with global navigation satellite system
(GNSS) receivers and can utilize the GNSS reference signals for synchronization.
The UAV-to-ground BS and the UAV-to-UAV links can be also utilized
to improve the accuracy of synchronization by adopting e.g. the precision
time protocol \cite[Ch. 8]{Fettweis11CoMP}.}. The data symbols intended for user $k$, denoted by $s_{k}$, are
modeled as Gaussian random variables with $s_{k}\sim\mathcal{CN}\left(0,1\right)$.
Let $\mathbf{w}_{l,k,t}\in\mathbb{C}^{M}$ be the beamforming vector
employed at UAV $l$ for sending $s_{k}$ at time $t$. Consequently,
the data signal received at user $k$ at time $t$ is given by
\begin{equation}
y_{k,t}=\sum\nolimits _{l\in\mathcal{L}}\mathbf{h}_{l,k,t}^{\mathrm{H}}\left(\sum\nolimits _{k\in\mathcal{K}}\mathbf{w}_{l,k,t}s_{k}\right)+z_{k,t},\label{eq:rx:signal}
\end{equation}
where $z_{k,t}\sim\mathcal{CN}\left(0,\sigma_{k}^{2}\right)$ is the
additive white Gaussian noise (AWGN) received at user $k$.

To enable dynamic cooperation among UAVs in \eqref{eq:rx:signal},
we require 
\begin{equation}
\left(1-q_{l,k}\right)\left\Vert \mathbf{w}_{l,k,t}\right\Vert =0,\quad l\in\mathcal{L},k\in\mathcal{K},t\in\mathcal{T},\label{eq:coop:topology}
\end{equation}
such that the beamformed radiation pattern of the UAVs' antenna array
is adapted to $\left\{ q_{l,k}\right\} $. In particular, if $q_{l,k}=0$,
we have $\mathbf{w}_{l,k,t}=\mathbf{0}$, $\forall t\in\mathcal{T}$,
and UAV $l$ does not transmit to user $k$; otherwise, $\mathbf{w}_{l,k,t}$
is unconstrained by \eqref{eq:coop:topology}. Furthermore, each user's
data symbols need to be conveyed to the cooperating UAVs in set $\mathcal{L}_{k}$
via wireless fronthauling. As spatial multiplexing is not beneficial
in low-rank LoS channels\footnote{Although we assume LoS fronthauling channels in this paper, the proposed
communication and optimization schemes are also applicable for other
fronthauling channel models. }\cite[Ch. 7.2.3]{Tse05Fundamentals}, the ground BS transmits only
a single data stream $s_{\mathrm{F},l}\sim\mathcal{CN}\left(0,1\right)$
to UAV $l\in\mathcal{L}$. Assume that the ground BS employs beamforming
vector $\mathbf{w}_{\mathrm{F},l,t}\in\mathbb{C}^{N}$ for transmitting
$s_{\mathrm{F},l}$ at time $t$. Consequently, the data signal received
at UAV $l\in\mathcal{L}$ at time $t$ during fronthauling is given
by 
\begin{equation}
\mathbf{y}_{\mathrm{F},l,t}=\mathbf{H}_{\mathrm{F},l,t}^{\mathrm{H}}\left(\sum\nolimits _{j\in\mathcal{L}}\mathbf{w}_{\mathrm{F},j,t}s_{\mathrm{F},j}\right)+\mathbf{z}_{\mathrm{F},l,t},\label{eq3}
\end{equation}
where $\mathbf{z}_{\mathrm{F},l,t}\sim\mathcal{CN}(\mathbf{0},\sigma_{\mathrm{F},l}^{2}\mathbf{I}_{M})$
is the AWGN. We note that beamforming is considered in \eqref{eq3}
for fronthauling to reap the power gains enabled by the multiple transmit
antennas at the ground BS. 

\vspace{-.4cm}

\subsection{Achievable Data Rate }

Assume that UAV $l$ employs the minimum mean squared error (MMSE)
beamforming for receiving $s_{\mathrm{F},l}$ \cite[Ch. 8.3]{Tse05Fundamentals}.
The achievable rate of UAV $l$ during wireless fronthauling is $R_{\mathrm{F},l,t}=\frac{1}{2}\log_{2}\left(1+\Gamma_{\mathrm{F},l}\right)$,
where $\Gamma_{\mathrm{F},l,t}$ is the signal-to-interference-plus-noise
ratio (SINR) given by 
\begin{equation}
\Gamma_{\mathrm{F},l,t}=\frac{A_{\mathrm{F},l}||\mathbf{G}_{\mathrm{F},l}^{\mathrm{H}}\mathbf{w}_{\mathrm{F},l,t}||^{2}/d_{\mathrm{F},l,t}^{\alpha_{\mathrm{F},l}}}{\sigma_{\mathrm{F},l}^{2}+\sum_{j\neq l}A_{\mathrm{F},l}||\mathbf{G}_{\mathrm{F},l}^{\mathrm{H}}\mathbf{w}_{\mathrm{F},j,t}||^{2}/d_{\mathrm{F},l,t}^{\alpha_{\mathrm{F},l}}}.\label{eq4-2}
\end{equation}
Moreover, the achievable rate of user $k\in\mathcal{K}$ is $R_{\mathrm{D},k,t}=\frac{1}{2}\log_{2}\left(1+\Gamma_{\mathrm{D},k,t}\right)$
and the SINR is given by 
\begin{equation}
\Gamma_{\mathrm{D},k,t}=\frac{\sum_{l\in\mathcal{L}}A_{l,k}|\mathbf{g}_{l,k}^{\mathrm{H}}\mathbf{w}_{l,k,t}|^{2}/d_{l,k,t}^{\alpha_{l,k}}}{\sigma_{k}^{2}\!+\!\sum_{l\in\mathcal{L}}\!\sum_{j\neq k}\!A_{l,k}|\mathbf{g}_{l,k}^{\mathrm{H}}\mathbf{w}_{l,j,t}|^{2}\!/\!d_{l,k,t}^{\alpha_{l,k}}},\label{eq5}
\end{equation}
provided that $R_{\mathrm{F},l,t}\ge\sum_{k\in\mathcal{K}}q_{l,k}R_{\mathrm{D},k,t}$.
The factor $\frac{1}{2}$ in the expressions for $R_{\mathrm{F},l,t}$
and $R_{\mathrm{D},k,t}$ is due to the time division between fronthauling
and data transmission. 

\vspace{-.4cm}

\section{\label{sec3}Problem Formulation}

Given the locations and the QoS requirements of the ground users,
in this section, joint optimization of the UAVs' navigation, cooperative
beamforming for data transmission, and beamforming for fronthauling
for maximization of the performance of the considered system is investigated.
This joint optimization is crucial as the UAVs' navigation and transmissions
simultaneously affect the signal and interference powers, and hence,
the achievable data rate of the users.

Let $\mathbf{d}_{l,0}$ be the initial location of UAV $l\in\mathcal{L}$.
The optimization space includes the UAVs' trajectories $\mathbf{d}\triangleq\left(\mathbf{d}_{l,t}\right)$,
i.e., the UAVs' positions in each time slot, and the cooperative transmission
policy $\mathbf{w}\triangleq\left(\mathbf{w}_{l,k,t},\mathbf{w}_{\mathrm{F},l,t},q_{k,l}\right)$.
We assume that a central controller located e.g. at the ground BS
is available for collecting the channel state information (CSI) $\mathbf{g}\triangleq\left(\mathbf{g}_{k},\mathbf{G}_{\mathrm{F},l}\right)$
and tracking the UAVs' positions $\mathbf{d}$. To minimize the UAVs'
power consumption while, at the same time, preventing the overloading
of the ground BS, the central controller computes the optimal trajectories
and beamformers with the objective to minimize the weighted sum of
the powers consumed by the UAVs and the ground BS while guaranteeing
the users' QoS requirements. The optimal decisions are fed back to
the UAVs and the ground BS for execution. The optimization problem
within block $\mathcal{T}$ is formulated as follows
\begin{alignat}{1}
\min_{\mathbf{w},\mathbf{d}}\;\; & \sum\nolimits _{t\in\mathcal{T}}f_{t}\left(\mathbf{w},\mathbf{d}\right)\label{eq:problem}\\
\mathrm{s.t.}\;\;\, & \textrm{\textrm{\ensuremath{\mathrm{\textrm{C1: }}}}}\sum\nolimits _{l\in\mathcal{L}}\left\Vert \mathbf{w}_{\mathrm{F},l,t}\right\Vert ^{2}\le P_{\mathrm{BS}}^{\max},\,t\in\mathcal{T}\nonumber \\
 & \textrm{C2: }\sum\nolimits _{k\in\mathcal{K}}\!\left\Vert \mathbf{w}_{l,k,t}\right\Vert ^{2}\!+\!P_{\mathrm{Nav},l,t}\!\le\!P_{l}^{\max},\,l\in\mathcal{L},t\in\mathcal{T}\nonumber \\
 & \textrm{C3: }q_{l,k}\in\left\{ 0,1\right\} ,\;l\in\mathcal{L},k\in\mathcal{K}\nonumber \\
 & \mathrm{\textrm{C4: }}\max_{t\in\mathcal{T}}\left\Vert \mathbf{w}_{l,k,t}\right\Vert ^{2}\le P_{l}^{\max}q_{l,k},\;l\in\mathcal{L},k\in\mathcal{K}\nonumber \\
 & \mathrm{\textrm{C5: }}\Gamma_{\mathrm{D},k,t}\ge\Gamma_{k}^{\min},\;k\in\mathcal{K},t\in\mathcal{T}\nonumber \\
 & \mathrm{\textrm{C6: }}\Gamma_{\mathrm{F},l,t}\ge2^{\sum\nolimits _{k\in\mathcal{K}}q_{l,k}R_{k}^{\min}}-1,\,l\in\mathcal{L},t\in\mathcal{T}\nonumber \\
 & \mathrm{\textrm{C7: }}\left\Vert \mathbf{d}_{l,t}-\mathbf{d}_{l,t-1}\right\Vert \le d^{\max},\;l\in\mathcal{L},t\in\mathcal{T}\nonumber \\
 & \mathrm{\textrm{C8: }}\left\Vert \mathbf{d}_{l,t}-\mathbf{d}_{j,t}\right\Vert \ge d^{\min},\;l,j\in\mathcal{L},\;l\neq j,t\in\mathcal{T}\nonumber \\
 & \mathrm{\textrm{C9: }}\mathbf{d}_{\mathrm{Nav}}^{\min}\preceq\mathbf{d}_{l,t}\preceq\mathbf{d}_{\mathrm{Nav}}^{\max},\;l\in\mathcal{L},t\in\mathcal{T},\nonumber 
\end{alignat}
where $f_{t}\left(\mathbf{w},\mathbf{d}\right)=\sum\nolimits _{l\in\mathcal{L}}\alpha_{l}(\sum\nolimits _{k\in\mathcal{K}}\left\Vert \mathbf{w}_{l,k,t}\right\Vert ^{2}+P_{\mathrm{Nav},l,t})+\alpha_{0}\sum\nolimits _{l\in\mathcal{L}}\left\Vert \mathbf{w}_{\mathrm{F},l,t}\right\Vert ^{2}$
is the weighted sum of the power consumptions of the UAVs and the
ground BS. The weights $\alpha_{l}$, $l\in\mathcal{L}$, and $\alpha_{0}$
assigned for UAV $l$ and the ground BS satisfy $\alpha_{l}\in[0,1]$
and $\alpha_{0}=1-\sum_{l\in\mathcal{L}}\alpha_{l}\in\left[0,1\right]$.
$P_{\mathrm{Nav},l,t}$ is the power consumed for hovering and repositioning
of UAV $l$ and is a function of the flight distance. In this paper,
we assume $P_{\mathrm{Nav},l,t}=c_{1}+c_{2}\left\Vert \mathbf{d}_{l,t}-\mathbf{d}_{l,t-1}\right\Vert $,
where constants $c_{1}$ and $c_{2}$ capture the power required for
keeping the UAV in the air and the power consumed for movement over
unit distance, respectively \cite{Seddon2001basic}.

In \eqref{eq:problem}, C1 constrains the maximum transmit power of
the ground BS to $P_{\mathrm{BS}}^{\max}$. C2 limits the maximum
power consumption of UAV $l$ to $P_{l}^{\max}$. C3 and C4 adjust
the cooperative UAV beamforming pattern for data transmission. We
note that C4 is an equivalent reformulation of \eqref{eq:coop:topology}
via the big-M technique \cite{Xiang16TWC:CoMP}: We have $\mathbf{w}_{l,k,t}=\mathbf{0}$
if $q_{l,k}=0$; otherwise, C4 ensures that the maximum power, $P_{l}^{\max}$,
of UAV $l$ is not exceeded. Moreover, as C4 is convex, it is more
convenient to deal with than \eqref{eq:coop:topology}. C5 and C6
limit the minimum instantaneous SINR/achievable rate for data transmission
and fronthauling, respectively. C5 and C6 together guarantee a minimum
instantaneous achievable rate of $R_{k}^{\min}=\frac{1}{2}\log_{2}\left(1+\Gamma_{k}^{\min}\right)$
{[}bps/Hz{]} for user $k$. Furthermore, C7 constrains the flight
range of UAV $l$ at time $t$ to be within an Euclidean ball of radius
$d^{\max}$ centered at its previous position, $\mathbf{d}_{l,t-1}$.
Herein, $d^{\max}$ depends on the UAVs' maximum flight speed and
the duration of a time slot. C8 ensures that any two UAVs are separated
by at least $d^{\min}$ for safe navigation. Finally, C9 specifies
the navigation zone of the UAVs with the boundaries defined by  $\mathbf{d}_{\mathrm{Nav}}^{\min}$
and $\mathbf{d}_{\mathrm{Nav}}^{\max}$. 

\vspace{-.4cm}

\section{Proposed Solution}

Problem \eqref{eq:problem} is a mixed-integer non-convex optimization
problem due to the binary variables $q_{l,k}$, which facilitate dynamic
UAV cooperation, and the non-convex constraints C5, C6, and C8, which
determine the UAVs' communication and navigation strategy.  This
type of problem is generally NP-hard and finding the global optimal
solution incurs an exponential-time computational complexity \cite{Xiang16TWC:CoMP}.
To balance between system performance and computational complexity,
in this section, we propose a low-complexity suboptimal algorithm
based on binary approximation and CCP to find a KKT solution. 

\vspace{-.4cm}

\subsection{Problem Transformation}

\subsubsection{Binary Approximation}

Recall that $q_{l,k}$ is a Dirac-like function of $\mathbf{w}_{l,k,t}$:
$q_{l,k}=0$ if and only if $\mathbf{w}_{l,k,t}=\mathbf{0}$ and $q_{l,k}=1$
otherwise. This motivates us to approximate $q_{l,k}$ using the following
family of functions, 
\begin{equation}
q_{l,k}\approx Q\left(\beta,\mathbf{w}_{l,k,t}\right)\triangleq1-\exp(-\beta\left\Vert \mathbf{w}_{l,k,t}\right\Vert ^{2}),\label{eq:approx}
\end{equation}
parametrized by $\beta\in\mathbb{R}_{+}$. $Q(\beta,\mathbf{w}_{l,k,t})$
has the following properties:
\begin{enumerate}
\item $Q(\beta,\mathbf{w}_{l,k,t})$ is zero for $\mathbf{w}_{l,k,t}=\mathbf{0}$
and has a value close to one for sufficiently large transmit power,
$\left\Vert \mathbf{w}_{l,k,t}\right\Vert ^{2}$. 
\item For large $\beta$, $Q(\beta,\mathbf{w}_{l,k,t})$ decreases to zero
near $\mathbf{w}_{l,k,t}=\mathbf{0}$ at a fast rate.
\item $Q\left(\beta,\mathbf{w}_{l,k,t}\right)$ is a differentiable quasiconvex
function of $\mathbf{w}_{l,k,t}$. That is, given $\theta\in\mathbb{R}$,
the sublevel sets, $\left\{ \mathbf{w}_{l,k,t}\in\mathbb{C}^{M\times1}\mid Q\left(\beta,\mathbf{w}_{l,k,t}\right)\le\theta\right\} $,
are convex.
\end{enumerate}
Properties 1) and 2) above imply that employing a large $\beta$ yields
an accurate approximation of the original $q_{l,k}$.  Substituting
\eqref{eq:approx} into problem \eqref{eq:problem}, C3 and C4 are
eliminated and the resulting optimization problem comprises only continuous
variables. Moreover, by reformulating the non-convex constraints explicitly
in DC form, problem \eqref{eq:problem} can be solved using conventional
convex optimization tools, as detailed subsequently.

\subsubsection{DC Reformulation}

First, we rewrite C5 and C6 as follows
\begin{alignat*}{1}
 & \mathrm{\textrm{C5: }}\sum\nolimits _{l\in\mathcal{L}}\tfrac{\gamma_{k}\left|\mathbf{g}_{l,k}^{\mathrm{H}}\mathbf{w}_{l,k,t}\right|^{2}-\sum_{j\in\mathcal{K}}\left|\mathbf{g}_{l,k}^{\mathrm{H}}\mathbf{w}_{l,j,t}\right|^{2}}{A_{l,k}^{-1}d_{l,k,t}^{\alpha_{l,k}}}\ge\sigma_{k}^{2},\;\forall k,\forall t\\
 & \mathrm{\textrm{C6: }}\tfrac{\left\Vert \mathbf{G}_{\mathrm{F},l}^{\mathrm{H}}\mathbf{w}_{\mathrm{F},l,t}\right\Vert ^{2}}{\sigma_{\mathrm{F},l}^{2}A_{\mathrm{F},l}^{-1}d_{\mathrm{F},l,t}^{\alpha_{\mathrm{F},l}}+\sum_{j\neq l}\left\Vert \mathbf{G}_{\mathrm{F},l}^{\mathrm{H}}\mathbf{w}_{\mathrm{F},j,t}\right\Vert ^{2}}\!\!\ge\!2^{\sum\nolimits _{k\in\mathcal{K}}q_{l,k}R_{k}^{\min}}\!\!\!-\!1,\forall t,
\end{alignat*}
where $\gamma_{k}\triangleq1+\tfrac{1}{\Gamma_{k}^{\min}}$. For given
$q_{l,k}$s, the positioning and beamforming variables $d_{\mathrm{F},l,t}$
and $\mathbf{w}_{\mathrm{F},j,t}$ in C6 are only loosely coupled,
as the fronthauling links have a common transmitter, i.e., the ground
BS; in contrast, for multiple UAVs, $d_{l,k,t}$ and $\mathbf{w}_{l,k,t}$
in C5 are tightly coupled such that the data transmission of one UAV
is affected by all other UAVs. By substituting \eqref{eq:approx}
and introducing slack variables $\boldsymbol{\tau}\triangleq\left(\tau_{l,k,t},\tau_{\mathrm{F},l},\tau_{q,l,k}\right)$,
we obtain an equivalent representation of C5 and C6 as follows
\begin{alignat*}{1}
 & \overline{\mathrm{\textrm{C5a}}}\textrm{: }\sum_{l\in\mathcal{L}}\tfrac{\gamma_{k}\left|\mathbf{g}_{l,k}^{\mathrm{H}}\mathbf{w}_{l,k,t}\right|^{2}}{\sigma_{k}^{2}\tau_{l,k,t}}\ge1+\sum_{l\in\mathcal{L}}\tfrac{\sum_{j\in\mathcal{K}}\left|\mathbf{g}_{l,k}^{\mathrm{H}}\mathbf{w}_{l,j,t}\right|^{2}}{\sigma_{k}^{2}\tau_{l,k,t}},\,\forall k,\forall t\\
 & \overline{\mathrm{\textrm{C5b}}}\textrm{: }\tau_{l,k,t}\ge A_{l,k}^{-1}d_{l,k,t}^{\alpha_{l,k}},\;\forall l,\forall k,\forall t\\
 & \overline{\mathrm{\textrm{C6a}}}\textrm{: }\tfrac{\left\Vert \mathbf{G}_{\mathrm{F},l}^{\mathrm{H}}\mathbf{w}_{\mathrm{F},l,t}\right\Vert ^{2}}{\sigma_{\mathrm{F},l}^{2}\tau_{\mathrm{F},l}}\ge A_{\mathrm{F},l}^{-1}d_{\mathrm{F},l,t}^{\alpha_{\mathrm{F},l}}+\tfrac{\sum_{j\neq l}\left\Vert \mathbf{G}_{\mathrm{F},l}^{\mathrm{H}}\mathbf{w}_{\mathrm{F},j,t}\right\Vert ^{2}}{\sigma_{\mathrm{F},l}^{2}},\forall l,\forall t\\
 & \overline{\mathrm{\textrm{C6b}}}\textrm{: }\log_{2}\left(1+\tau_{\mathrm{F},l}\right)\ge\sum\nolimits _{k\in\mathcal{K}}R_{k}^{\min}\tau_{q,l,k},\;\forall l\\
 & \overline{\mathrm{\textrm{C6c}}}\textrm{: }\tau_{q,l,k}\!\ge\!1\!-\!e^{-\beta\left\Vert \mathbf{w}_{l,k}\right\Vert ^{2}}\!\!\!\!\iff\!\!\!\beta\left\Vert \mathbf{w}_{l,k}\right\Vert ^{2}\!\le\!-\!\ln\left(1\!-\!\tau_{q,l,k}\right).
\end{alignat*}
Here, $\overline{\mathrm{\textrm{C5b}}}$ and $\overline{\mathrm{\textrm{C6b}}}$
are convex constraints. $\overline{\mathrm{\textrm{C5a}}}$, $\overline{\mathrm{\textrm{C6a}}}$,
and $\overline{\mathrm{\textrm{C6c}}}$ are DC constraints, where
both sides of each inequality are convex functions \cite{Boyd2004Convex}.
Moreover, C8 is already in DC form.

Next,  by defining $\mathbf{x}\triangleq(\mathbf{w},\mathbf{d},\boldsymbol{\tau})$,
problem \eqref{eq:problem} can be solved approximately by solving,

\vspace{-0.5cm}

\begin{alignat}{1}
\min_{\mathbf{x}}\;\; & f_{0}\left(\mathbf{x}\right)\triangleq\sum\nolimits _{t\in\mathcal{T}}f_{t}\left(\mathbf{x}\right)\label{eq:approx-problem}\\
\mathrm{s.t.}\;\;\, & \mathbf{x}\in\mathcal{X}\triangleq\left\{ \mathbf{x}\mid\textrm{\textrm{\ensuremath{\mathrm{\textrm{C1, C2, \ensuremath{\overline{\mathrm{\textrm{C5b}}}}, \ensuremath{\overline{\mathrm{\textrm{C6b}}}\textrm{,}} C7}}}}}\right\} \nonumber \\
 & \mathbf{f}_{1}(\mathbf{x})-\mathbf{f}_{2}(\mathbf{x})\preceq\mathbf{0},\nonumber 
\end{alignat}
where $\mathbf{f}_{1}(\cdot)$ and $\mathbf{f}_{2}(\cdot)$ are convex
functions representing the DC constraints $\overline{\mathrm{\textrm{C5a}}}$,
$\overline{\mathrm{\textrm{C6a}}}$, $\overline{\mathrm{\textrm{C6b}}}$,
and C8. Problems \eqref{eq:approx-problem} and \eqref{eq:problem}
are equivalent, in the sense that both problems have the same optimal
value and optimal solution, for $\beta\to\infty$.

\vspace{-.4cm}

\subsection{Proposed Iterative Algorithm}

Problem \eqref{eq:approx-problem} is a reverse convex program \cite{Lipp16CCP},
which optimizes a convex objective function over a feasible set formed
by both DC and convex constraints. We solve problem \eqref{eq:approx-problem}
using an iterative approximation procedure. Let $m$ be the iteration
index. Assume for the moment that $\mathbf{x}_{(m-1)}$ is a given
feasible point of problem \eqref{eq:approx-problem}, e.g., obtained
in iteration $m-1$. In iteration $m$, we approximate $\mathbf{f}_{2}(\mathbf{x})$
using the first-order Taylor approximation at $\mathbf{x}_{(m-1)}$
\begin{equation}
\widetilde{\mathbf{f}}_{2}(\mathbf{x}\mid\mathbf{x}_{(m-1)})\!\triangleq\!\mathbf{f}_{2}(\mathbf{x}_{(m-1)})\!+\!\nabla\mathbf{f}_{2}(\mathbf{x}_{(m-1)})^{\mathrm{T}}(\mathbf{x}-\mathbf{x}_{(m-1)}),\label{eq:f2}
\end{equation}
with $\widetilde{\mathbf{f}}_{2}(\mathbf{x}_{(m-1)}\mid\mathbf{x}_{(m-1)})=\mathbf{f}_{2}(\mathbf{x}_{(m-1)})$.
As $\mathbf{f}_{1}(\mathbf{x})-\widetilde{\mathbf{f}}_{2}(\mathbf{x}\mid\mathbf{x}_{(m-1)})$
is convex, convex problem

\vspace{-0.5cm}

\begin{alignat}{1}
\min_{\mathbf{x}}\;\; & f_{0}\left(\mathbf{x}\right)\label{eq:ccp}\\
\mathrm{s.t.}\;\;\, & \mathbf{x}\in\mathcal{X},\;\mathbf{f}_{1}(\mathbf{x})-\widetilde{\mathbf{f}}_{2}(\mathbf{x}\mid\mathbf{x}_{(m-1)})\preceq\mathbf{0},\nonumber 
\end{alignat}
can be solved optimally using standard solvers such as CVX \cite{Boyd2004Convex}.
Now, denote the optimal solution of \eqref{eq:ccp} by $\mathbf{x}_{(m)}$.
As $\mathbf{f}_{2}(\mathbf{x})$ is convex, we have $\mathbf{f}_{2}(\mathbf{x})\succeq\widetilde{\mathbf{f}}_{2}(\mathbf{x}\mid\mathbf{x}_{(m-1)})$
and $\mathbf{f}_{1}(\mathbf{x})-\mathbf{f}_{2}(\mathbf{x})\preceq\mathbf{f}_{1}(\mathbf{x})-\widetilde{\mathbf{f}}_{2}(\mathbf{x}\mid\mathbf{x}_{(m-1)})$,
$\forall\mathbf{x}$. We can further show that \cite{Lipp16CCP}
\begin{enumerate}
\item $\mathbf{f}_{1}(\mathbf{x}_{(m)})-\mathbf{f}_{2}(\mathbf{x}_{(m)})\preceq\mathbf{0}$,
i.e., $\mathbf{x}_{(m)}$ is a feasible point for problem \eqref{eq:approx-problem}
as well,
\item $f_{0}(\mathbf{x}_{(m)})\ge f_{0}^{*}$, i.e., $f_{0}(\mathbf{x}_{(m)})$
gives an upper bound for the optimal value of problem \eqref{eq:approx-problem},
$f_{0}^{*}$, and 
\item $f_{0}(\mathbf{x}_{(m)})\le f_{0}(\mathbf{x}_{(m-1)})$ as $\mathbf{x}_{(m-1)}$
is feasible (though possibly not optimal) for problem \eqref{eq:ccp}.
\end{enumerate}
Therefore, by successively employing \eqref{eq:f2} and solving the
resulting problem \eqref{eq:ccp}, we obtain a non-increasing sequence
of solutions $\left\{ \mathbf{x}_{(m)}\right\} $ of problem \eqref{eq:approx-problem}.
The iterative process is summarized in Algorithm~\ref{alg1}. Similar
to \cite{Lipp16CCP}, we can show that Algorithm~\ref{alg1} converges
to a KKT point of problem \eqref{eq:approx-problem} (and problem
\eqref{eq:problem} for a large $\beta$) after a sufficiently large
number of iterations. Note that, in line~\ref{line4} of Algorithm~\ref{alg1},
$\mathbf{x}_{(m)}$ can be computed within polynomial time. Therefore,
the overall computational complexity of Algorithm~\ref{alg1} grows
only polynomially with the size of problems \eqref{eq:approx-problem}
and \eqref{eq:problem}.

\begin{algorithm}[t] 
\protect\caption{Proposed  Algorithm for Solving \eqref{eq:approx-problem} and \eqref{eq:problem}}
\label{alg1}  
\small{ \begin{algorithmic}[1] 

\STATE  \textbf{initialization}: 
{Set}  maximum number of iterations, $N_{\mathrm{it}}$, and tolerance, $\epsilon$; $m \leftarrow 1$; 

\STATE Find a feasible point $\mathbf{x}_{(0)} \triangleq (\mathbf{w}_{(0)},\mathbf{d}_{(0)},\boldsymbol{\tau}_{(0)}) $ by solving problem \eqref{eq:problem2};

\REPEAT  
\STATE  Solve problem \eqref{eq:ccp} and obtain the optimal solution $\mathbf{x}_{(m)}$; \label{line4}
\STATE  Compute: $\mathsf{error} \leftarrow f_0(\mathbf{x}_{(m-1)}) - f_0(\mathbf{x}_{(m)})$
\STATE  Update: $m \leftarrow m+1 $;
\UNTIL {$\mathsf{error} \le \epsilon$ or $m > N_{\mathrm{it}}$.} \label{alg1:line17}  \end{algorithmic}  } 
\end{algorithm}

Two remarks regarding Algorithm~\ref{alg1} are in order. First, as
$\mathbf{f}_{2}(\cdot)$ involves quadratic-over-linear functions
of complex-valued variable $\mathbf{w}$, we have to determine a real-valued
lower-bound function $\widetilde{\mathbf{f}}_{2}(\cdot\mid\cdot)$
for $\mathbf{f}_{2}(\cdot)$ in \eqref{eq:f2}, which is given in
Lemma \ref{lem1}. 

\vspace{-.2cm}
\begin{lem}
\emph{\label{lem1}The quadratic-over-linear function $f\left(\mathbf{w},\tau\right)\triangleq\unitfrac{\left\Vert \mathbf{G}^{H}\mathbf{w}\right\Vert ^{2}}{\tau}$,
which is defined on $\mathbb{C}^{N}\times\mathbb{R}_{+}\to\mathbb{R}_{+}$
for given $\mathbf{G}\in\mathbb{C}^{N\times M}$, is lower bounded
at $\left(\mathbf{w}_{(0)},\tau_{(0)}\right)$ as
\begin{equation}
f\left(\mathbf{w},\tau\right)\!\ge\!f\left(\mathbf{w}_{(0)},\tau_{(0)}\right)+\tfrac{\mathbf{\widetilde{w}}_{(0)}^{\mathrm{T}}\widetilde{\mathbf{G}}\widetilde{\mathbf{G}}^{\mathrm{T}}}{\tau_{(0)}}\!\left(2\widetilde{\mathbf{w}}\!-\!\tfrac{\tau+\tau_{(0)}}{\tau_{(0)}}\mathbf{\widetilde{w}}_{(0)}\right),\label{eq9}
\end{equation}
where $\widetilde{\mathbf{G}}\triangleq\left[\!\!\begin{array}{cc}
\Re\left\{ \mathbf{G}\right\}  & -\Im\left\{ \mathbf{G}\right\} \\
\Im\left\{ \mathbf{G}\right\}  & \Re\left\{ \mathbf{G}\right\} 
\end{array}\!\!\right]\in\mathbb{\mathbb{R}}^{2N\times2M}$, }$\widetilde{\mathbf{w}}\triangleq\left[\!\!\begin{array}{c}
\Re\left\{ \mathbf{w}\right\} \\
\Im\left\{ \mathbf{w}\right\} 
\end{array}\!\!\right]\in\mathbb{\mathbb{R}}^{2N}$\emph{, and} $\mathbf{\widetilde{w}}_{(0)}=\left[\!\!\begin{array}{c}
\Re\left\{ \mathbf{w}_{(0)}\right\} \\
\Im\left\{ \mathbf{w}_{(0)}\right\} 
\end{array}\!\!\right]\in\mathbb{\mathbb{R}}^{2N}$.
\end{lem}
\begin{IEEEproof}
The result holds due to $f\left(\mathbf{w},\tau\right)=\widetilde{f}\left(\widetilde{\mathbf{w}},\tau\right)\triangleq||\widetilde{\mathbf{G}}^{\mathrm{T}}\widetilde{\mathbf{w}}||^{2}/\tau$,
where $\widetilde{f}\left(\widetilde{\mathbf{w}},\tau\right)$ is
a jointly convex function of $\left(\widetilde{\mathbf{w}},\tau\right)$
defined on\emph{ $\mathbb{R}^{2N}\times\mathbb{R}_{+}\to\mathbb{R}_{+}$.}
However, the detailed proof is ignored for saving space.
\end{IEEEproof}
Second, Algorithm~\ref{alg1} requires the starting point $\mathbf{x}_{(0)}=(\mathbf{w}_{(0)},\mathbf{d}_{(0)},\boldsymbol{\tau}_{(0)})$
to be feasible for problem \eqref{eq:problem}. To this end, we first
define trajectories $\mathbf{d}_{(0)}$ and cooperation decisions
$\mathbf{q}_{(0)}$ according to C7, C8, and C9. The navigation power
$P_{\mathrm{Nav},l,t}$ is determined by $\mathbf{d}_{(0)}$. Then,
let $\mathbf{W}_{\mathrm{F},l,t}=\mathbf{w}_{\mathrm{F},l,t}\mathbf{w}_{\mathrm{F},l,t}^{\mathrm{H}}$
and $\mathbf{W}_{l,k,t}=\mathbf{w}_{l,k,t}\mathbf{w}_{l,k,t}^{\text{H}}$.
By fixing $\mathbf{d}=\mathbf{d}_{(0)}$ and $\mathbf{q}=\mathbf{q}_{(0)}$
in \eqref{eq:problem}, the following semi-definite optimization problem
is obtained from \eqref{eq:problem}, 

\vspace{-0.5cm}

\begin{alignat}{1}
\min\; & \sum\nolimits _{l\in\mathcal{L}}(\alpha_{0}\mathrm{tr}\left(\mathbf{W}_{\mathrm{F},l,t}\right)+\sum\nolimits _{k\in\mathcal{K}}\alpha_{l}\mathrm{tr}\left(\mathbf{W}_{l,k,t}\right))\label{eq:problem2}\\
\mathrm{s.t.}\;\, & \textrm{\textrm{\ensuremath{\mathrm{\textrm{C1: }}}}}\sum\nolimits _{l\in\mathcal{L}}\mathrm{tr}\left(\mathbf{W}_{\mathrm{F},l,t}\right)\le P_{\mathrm{BS}}^{\max},\;\forall t\nonumber \\
 & \textrm{C2: }\sum\nolimits _{k\in\mathcal{K}}\!\mathrm{tr}\left(\mathbf{W}_{l,k,t}\right)\!\le\!P_{l}^{\max}-P_{\mathrm{Nav},l,t},\,\forall l,\forall t\nonumber \\
 & \mathrm{\textrm{C4: }}\max_{t\in\mathcal{T}}\mathrm{tr}\left(\mathbf{W}_{l,k}\right)\le P_{l}^{\max}q_{l,k},\;\forall l,\forall k\nonumber \\
 & \mathrm{\textrm{C5: }}\sum\nolimits _{l\in\mathcal{L}}\tfrac{\mathrm{tr}\left(\left(\gamma_{k}\mathbf{W}_{l,k,t}-\sum_{j\in\mathcal{K}}\mathbf{W}_{l,j,t}\right)\mathbf{g}_{l,k}\mathbf{g}_{l,k}^{\mathrm{H}}\right)}{A_{l,k}d_{l,k,t}^{\alpha_{l,k}}}\ge\sigma_{k}^{2}\nonumber \\
 & \mathrm{\textrm{C6: }}\tfrac{\mathrm{tr}\left(\left(\gamma_{\mathrm{F},l}\mathbf{W}_{\mathrm{F},l,t}-\sum_{j\in\mathcal{L}}\mathbf{W}_{\mathrm{F},j,t}\right)\mathbf{G}_{\mathrm{F},l}\mathbf{G}_{\mathrm{F},l}^{\mathrm{H}}\right)}{A_{\mathrm{F},l}d_{\mathrm{F},l,t}^{\alpha_{\mathrm{F},l}}}\!\ge\!\sigma_{\mathrm{F},l}^{2}\nonumber \\
 & \mathrm{\textrm{C10: }}\mathbf{W}_{\mathrm{F},l,t}\succeq\mathbf{0},\quad\mathbf{W}_{l,k,t}\succeq\mathbf{0},\;\forall l,\forall k,\forall t\nonumber \\
 & \mathrm{\textrm{C11: }}\mathrm{rank}(\mathbf{W}_{\mathrm{F},l,t})=\mathrm{rank}(\mathbf{W}_{l,k,t})=1,\;\forall l,\forall k,\forall t,\nonumber 
\end{alignat}
where $\gamma_{\mathrm{F},l}\triangleq1+\frac{1}{2^{\sum\nolimits _{k\in\mathcal{K}}q_{l,k}R_{k}^{\min}}\!\!-\!1}$.
Problem \eqref{eq:problem2} is solved using semi-definite relaxation,
i.e., by dropping the rank constraint C11. We can show, similar to
\cite[Theorem 1]{Xiang16TWC:CoMP}, that the obtained solutions, denoted
by $\mathbf{W}_{\mathrm{F},l,t}^{*}$ and $\mathbf{W}_{l,k,t}^{*}$,
both have rank one. Hence, $\mathbf{W}_{\mathrm{F},l,t}^{*}$ and
$\mathbf{W}_{l,k,t}^{*}$ are the optimal solutions of problem \eqref{eq:problem2}.\textcolor{red}{{}
}Consequently,\textcolor{red}{{} }we obtain the beamforming vectors
$\mathbf{w}_{(0)}$ as the principal eigenvectors of $\mathbf{W}_{\mathrm{F},l,t}^{*}$
and $\mathbf{W}_{l,k,t}^{*}$. Finally, based on $\mathbf{d}_{(0)}$
and $\mathbf{w}_{(0)}$, $\boldsymbol{\tau}_{(0)}$ is readily available
from \eqref{eq:approx-problem}.

\vspace{-.4cm}

\section{Performance Evaluation}

In this section, we evaluate the performance of the proposed dynamic
UAV cooperation scheme in an aerial network as shown in Figure \ref{fig1},
where $K=4$ ground users are randomly distributed within a ring with
inner radius $R_{1}=0.5$ km and outer radius $R_{2}=1$ km. We assume
that the disk is centered at the origin $O$. To serve the users,
$L=4$ UAVs are deployed within a cylindrical navigation space of
radius $R_{2}$, minimum height $50$ m, and maximum height $100$
m above the disk. The initial positions of the UAVs are randomly selected
within the defined navigation space. The ground BS located at the
origin $O$ provides fronthauling for the UAVs.  For simulating the
air-to-ground channels, the path losses are set according to the 3GPP
``Macro + Outdoor Relay'' scenario \cite{3GPP:TR36814} and the
channel fading is Rician distributed with Rice factor $-3$ dB. 
The other relevant system parameters are given in Table \ref{tab1}.
Each simulation is performed for $B=30$ time blocks, where in each
time block we minimize the total power consumption by solving problem
\eqref{eq:problem} using weights $\alpha_{0}=\alpha_{l}=\frac{1}{L+1}$,
$l\in\mathcal{L}$.

\begin{table}
\caption{\label{tab1}Simulation parameters}

\vspace{-.2cm}
\renewcommand{\arraystretch}{1.1}\centering\footnotesize

\begin{tabular}{|c|c|}
\hline 
Parameters & Settings\tabularnewline
\hline 
\hline 
System bandwidth & $2$ MHz\tabularnewline
\hline 
Duration of time slot & $0.2$ s\tabularnewline
\hline 
Number of time slots & $T=50$\tabularnewline
\hline 
Number of antennas & $N=12$, $M=2$\tabularnewline
\hline 
Transmit power  & $P_{\mathrm{BS}}^{\max}=46$ dBm, $P_{l}^{\max}=40$ dBm\tabularnewline
\hline 
Navigation power & $c_{1}=0$ dBm, $c_{2}=20$ dBm/m\tabularnewline
\hline 
Antenna height & $d_{k,z}^{\mathrm{UE}}=0$ m, $d_{z}^{\mathrm{BS}}=25$ m\tabularnewline
\hline 
Noise power spectral density & $-174$ dBm/Hz\tabularnewline
\hline 
Max. flying speed of UAVs & $10$ m/s\tabularnewline
\hline 
Min. data rate for users & $R_{k}^{\min}=0.8$ Mbps\tabularnewline
\hline 
Safety distance for UAVs & $d^{\min}=10$ m\tabularnewline
\hline 
\end{tabular}

\vspace{-.4cm}
\end{table}

For comparison, the following schemes are considered as baselines:
\begin{itemize}
\item \emph{Baseline Scheme 1 (Coordinated beamforming):} Each user is randomly
associated with one of the UAVs and each UAV serves at most $\min\left(M,K\right)$
users.
\item \emph{Baseline Scheme 2 (Fixed cooperation):} Each user is randomly
associated with at least one UAV such that each UAV serves $\min\left(M,K\right)$
users. For Baselines 1 and 2, $(\mathbf{w},\mathbf{d})$ is optimized
using Algorithm~\ref{alg1} with $\mathbf{q}$ fixed accordingly. 
\item \emph{Baseline Scheme 3 (Hovering)}: All UAVs keep hovering at their
initial positions, $\mathbf{d}_{0}$. 
\item \emph{Baseline Scheme 4 (Navigating along fixed trajectories)}: Each
UAV flies horizontally at a given speed to reach the boundary of the
navigation space at time $BT$. Each UAV flies along the path of shortest
length. For Baselines 3 and 4, $(\mathbf{w},\mathbf{q})$ is optimized
using Algorithm~\ref{alg1} with $\mathbf{d}$ fixed accordingly. 
\end{itemize}

\begin{figure}[t]
\centering\includegraphics[scale=0.52]{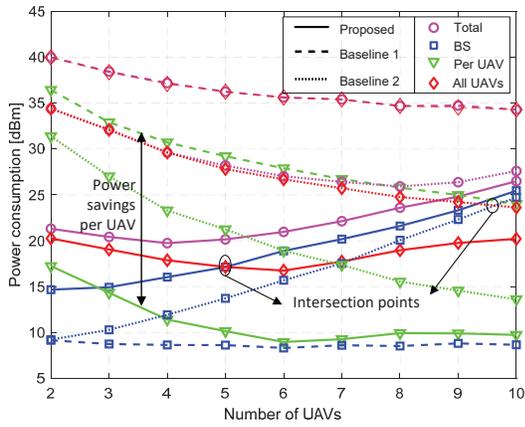}

\vspace{-.2cm}

\caption{\label{fig3}Power consumption versus number of UAVs, $L$.}

\vspace{-.2cm}
\end{figure}

Figure \ref{fig3} shows the power consumption of the considered schemes
as a function of the number of UAVs, $L$, where `BS', `all UAVs',
and `total' denote the power consumptions of the ground BS for fronthauling,
the power consumption of the UAVs for navigation and data transmission,
and the total power consumption, respectively. Moreover, `per UAV'
denotes the average power consumed per UAV for navigation and data
transmission. From Figure \ref{fig3} we observe that, as expected,
Baseline Scheme 1 provides an upper bound for the system's total power
consumption, due to power-inefficient data transmission among the
UAVs. However, with Baseline Scheme 1, the ground BS consumes less
transmit power compared to the other schemes for all considered values
of $L$, as each user's data needs to be delivered to only one UAV
via fronthauling. 

Compared with Baseline Scheme 1, Baseline Scheme 2 and the proposed
scheme significantly reduce the power required for data transmission
and navigation by enabling cooperative transmission among the UAVs
and exploiting the resulting large virtual antenna array. However,
as UAV cooperation requires a high data rate for the fronthauling
links, Baseline Scheme 2 and the proposed scheme require a higher
transmit power for the ground BS than Baseline Scheme 1. Furthermore,
with Baseline Scheme 2 and the proposed scheme, the power consumptions
of the ground BS even exceeds that of the UAVs for large $L$, where
the intersection points are also shown in the figure. This fact, along
with the increased power required to keep the UAVs in the air for
larger $L$, leads to increased total power consumption. This result
reveals an intricate trade-off between the power consumption for fronthauling,
data transmission, and navigation in cooperative multi-UAV systems,
whereby $L$ has to be optimized for minimization of the system's
total power consumption. Nevertheless, by optimizing the cooperation
decisions $\mathbf{q}$, the proposed scheme significantly reduces
the average power consumption per UAV and the system's total power
consumption compared to the baseline schemes despite the SWAP limitations,
at the expense of a small increase of the ground BS's transmit power.
For example, compared with Baseline Scheme 2, the average power consumed
per UAV with the proposed scheme reduces by more than $10$ dB for
$L\le6$, whereas the ground BS's transmit power is increased by less
than $6$ dB. 

\begin{figure}[t]
\centering\includegraphics[scale=0.54]{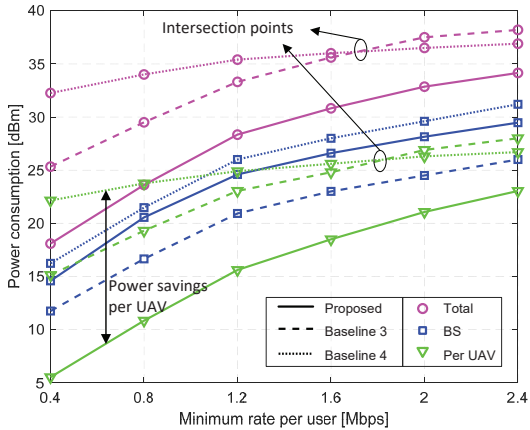}

\vspace{-.2cm}\caption{\label{fig4}Power consumption versus users' minimum required rate,
$R_{k}^{\min}$. }

\vspace{-.5cm}
\end{figure}

Figure \ref{fig4} illustrates the power consumption of the considered
schemes as a function of the minimum rate achievable at the users,
$R_{k}^{\min}$. From Figure \ref{fig4} we observe that the power
consumptions of both the UAVs and the ground BS increase monotonically
with $R_{k}^{\min}$, as more transmit power is needed to simultaneously
increase the rates for data transmission and fronthauling, cf. C5
and C6. Moreover, by optimizing the trajectories of the cooperating
UAVs, the proposed scheme significantly reduces the power consumption
per UAV for all considered $R_{k}^{\min}$s compared to the baseline
schemes. To gain insight regarding the importance of optimal trajectory
design, we note an interesting trade-off between the power consumptions
for navigation and communication (including fronthauling and data
transmission), which is revealed by Baseline Schemes 3 and 4, cf.
the intersection points shown in the figure. In particular, for Baseline
Scheme 3, the transmit power required by the ground BS for fronthauling
is low as the UAVs hover close to the ground BS, whereas the UAVs
may need a large power for data transmission, particularly when $R_{k}^{\min}$
is large, as they are far away from the users. In contrast, by having
the UAVs fly close to the users (away from the ground BS), which leads
to an increased navigation power, Baseline Scheme 4 reduces the power
consumption required for data transmission when $R_{k}^{\min}$ is
large, and the transmit power for fronthauling increases only slightly.
Therefore, when $R_{k}^{\min}$ is large, flying the UAVs close to
the users is preferable for lowing the UAVs' power consumed in data
transmission. On the other hand, when $R_{k}^{\min}$ is small, hovering
the UAVs close to the ground BS is preferred for lowering the power
consumption in navigation and fronthauling. 

\vspace{-.4cm}

\section{Conclusions}

In this paper, dynamic multi-UAV cooperation was investigated for enabling power-efficient aerial communications. Thereby,  the UAVs are intelligently selected for cooperatively serving the ground users
and the resulting large virtual array of moving antennas is exploited
to reduce the power consumptions of UAV navigation and communication.
The UAVs' trajectories and cooperative beamforming were jointly designed
by solving a mixed-integer non-convex optimization problem. As the
problem is NP-hard, a low-complexity algorithm exploiting the underlying
DC program structure was developed for finding a suboptimal solution.
Simulation results revealed interesting trade-offs between the powers
required for fronthauling, data transmission, and navigation in cooperative
multi-UAV systems. Moreover, the proposed dynamic multi-UAV cooperation
scheme can significantly lower the power consumption per UAV while
guaranteeing the users' QoS requirements, and hence, provides a promising
approach to enhance aerial communications.

\vspace{-.4cm}

\section*{Acknowledgment}

The authors are supported by the ERC AGNOSTIC project (grant R-AGR-3283)
and the FNR CORE projects 5G-Sky (C19/IS/13713801) and ROSETTA (11632107). 

\vspace{-.4cm}

\bibliographystyle{IEEEtran}
\bibliography{IEEEabrv,UAV,OptimizationRefs}

\end{document}